\documentclass[aps,prd,preprintnumbers,superscriptaddress,nofootinbib,showpacs,twocolumn]{revtex4}%
\usepackage{amsmath,amssymb,bm,float,mathrsfs,tabularx}
\usepackage[dvipdfmx]{graphicx}
\usepackage{dcolumn}
\usepackage{color}
\usepackage{hyperref}
\usepackage{comment}

  \newcommand{\beq}{\begin{equation}}
  \newcommand{\eeq}{\end{equation}}
  \newcommand{\al}[1]{\begin{align} #1 \end{align}}
  \newcommand{\bi}{\begin{itemize}}
  \newcommand{\ei}{\end{itemize}}
  \def\dd{\mathrm{d}}

  \def\pd{\partial}

\begin{document}

\title{Probing higher-order primordial non-Gaussianity with galaxy surveys}


\author{Daisuke Yamauchi}
\email[Email: ]{yamauchi"at"jindai.jp}
\affiliation{
Research Center for the Early Universe, Graduate School of Science, 
The University of Tokyo, Bunkyo-ku, Tokyo 113-0033, Japan
}
\affiliation{
Faculty of Engineering, Kanagawa University, Kanagawa-ku, Yokohama-shi, 
Kanagawa, 221-8686, Japan
}

\author{Keitaro Takahashi}
\affiliation{
Faculty of Science, Kumamoto University, 2-39-1 Kurokami, Kumamoto 860-8555, Japan
}

\begin{abstract}
With a radio continuum galaxy survey by the Square Kilometre Array (SKA), a photometric galaxy
survey by Euclid and their combination, we forecast future constraints on primordial non-Gaussianity.
We focus on the potential impact of local-type higher-order nonlinear parameters
on the parameter estimation and particularly the confirmation of the inflationary consistency inequality.
Nonstandard inflationary models, such as multifield models, 
introduce the scale-dependent stochastic clustering of galaxies on large scales,
which is a unique probe of mechanism for generating primordial density fluctuations.
Our Fisher matrix analysis indicates that 
a deep and wide survey provided by SKA is more advantageous to constrain $\tau_{\rm NL}$, while Euclid has a strong constraining power for $f_{\rm NL}$ due to the redshift information, suggesting that the joint analysis between them is quite essential to break the degeneracy between the nonlinear parameters.
The combination of the full SKA and Euclid will achieve the precision level needed 
to confirm the consistency inequality even for $f_{\rm NL}\approx 1.5$ and $\tau_{\rm NL}\approx 17$, though it is still hard for a single survey to confirm it when $f_{\rm NL}\lesssim 2.7$.

\end{abstract}

\preprint{RESCEU-25/15}

\maketitle

\section{Introduction} 

Observing the large-scale distribution of matter provides us the rich information 
about not only the late-time evolution but also the primordial nature of the Universe. 
Among various cosmological parameters characterizing the primordial Universe, we are particularly
interested in possible departures from a purely Gaussian distribution of primordial density fluctuations 
called primordial non-Gaussianity (PNG), which is one of the most powerful tests of inflation and 
more generally a key to understanding the extreme high-energy physics.

One of the major theoretical discoveries is that {\it all} inflationary models predict
the presence of the consistency relation between the parameters characterizing PNG.
For the simplest scenarios, if there is the nonvanishing local-form bispectrum, 
the trispectrum must necessarily exist with
$\tau_{\rm NL}=((6/5)f_{\rm NL})^2$, where $f_{\rm NL}$ and $\tau_{\rm NL}$
are the so-called local-type nonlinear parameters characterizing the amplitude of 
the primordial bispectrum and trispectrum.
Even in a general situation, one can show that there is a universal relation, 
$\tau_{\rm NL}\geq ((6/5)f_{\rm NL})^2$, which is often refereed to as the 
Suyama-Yamaguchi inequality~\cite{Suyama:2007bg,Suyama:2010uj} 
(see also \cite{Sugiyama:2011jt,Rodriguez:2013cj}).
The primordial trispectrum is also usually characterized by another nonlinear parameter, $g_{\rm NL}$, 
which corresponds to the strength of the intrinsic cubic nonlinearities of primordial fluctuations.
Hence a detection of the higher-order PNG and the confirmation of the inequality 
would indicate the presence of more complicated dynamics, e.g., multifield inflationary models, 
in the primordial Universe and should be, thus, the target in future experiments (see also \cite{Biagetti:2012xy}).

The current limits on these parameters have been obtained from cosmic microwave background (CMB)
anisotropies: $f_{\rm NL}^{\rm CMB}=0.8\pm 5.0$ and $g_{\rm NL}^{\rm CMB}=\left( -9.0\pm 7.7\right)\times 10^4$
at $1\sigma$ statistical significance~\cite{Ade:2015ava} and 
$\tau_{\rm NL}^{\rm CMB}<2800$ ($95\%$ C.L.)~\cite{Ade:2013ydc}.
However, CMB measurements are already close to being cosmic-variance limited.
A complementary way to access PNG is to measure its impact on a large-scale structure.
For the standard Gaussian initial conditions, the halo bias is often assumed to be linear, deterministic, and scale independent. 
It was found that the non-Gaussianity in primordial fluctuations effectively introduces 
a {\it scale-dependent} clustering of galaxies on large scales~\cite{Dalal:2007cu,Matarrese:2008nc}.
In this paper, we focus on scale-dependent {\it stochastic} halo bias.
If $\tau_{\rm NL}>((6/5)f_{\rm NL})^2$, the halo density contrast $\delta_{\rm h}$ 
is not $100\%$ correlated to the matter linear density field $\delta$ even in the absence of the shot noise~\cite{Smith:2010gx,Baumann:2012bc,Tseliakhovich:2010kf}.
Namely, the simple bias relation $\delta_{\rm h}=b_{\rm h}\delta$ should be modified due to the stochasticity.
The halo-halo power spectrum between $b$- and $b^\prime$-mass bins, $P_{\rm hh}^{(bb^\prime )}$, cannot
be expressed as the products of the linear bias defined by $b_{{\rm h}}^{(b)}\equiv P_{\rm mh}^{(b)}/P_{\rm mm}$ 
and the linear matter density power spectrum $P_{\rm mm}$. 
Formally this means $P_{\rm hh}^{(bb^\prime )}\geq b_{{\rm h}}^{(b)}b_{{\rm h}}^{(b^\prime )}P_{\rm mm}$.
This provides the unique opportunity of large-scale scale-dependent {\it stochastic} bias as a probe of the primordial Universe
associated with complicated dynamics.

A promising way to explore PNG in future large-scale structure surveys
is to reduce the cosmic-variance noise with the so-called multitracer technique~\cite{Seljak:2008xr,Hamaus:2011dq},
which allows us to measure the scale-dependent bias very accurately.
Since using the multitracer technique we can constrain the PNG in the halo bias
without suffering from the cosmic-variance noise, the clustering analysis in this case is expected to be limited 
only due to the contributions from the shot noise. 
In order to reduce the remaining noise contribution, the two-dimensional map of the large-scale structure
projected along the line of sight is considered, while the redshift information of galaxies would be lost by the projection. 
Although the multitracer technique is truly effective for three dimensional statistics,
in this paper we will focus on two-dimensional one as a simple extension of our previous work~\cite{Yamauchi:2014ioa}.
Even when we use the projected density contrasts as tracers, the multitracer technique is expected to be still effective.
Indeed Ref.~\cite{Yamauchi:2014ioa} shows that the constraining power on $f_{\rm NL}$
drastically improves even when splitting samples into two tracers and increases with the number of tracers.
In this paper we extend the multitracer technique to include the effects of 
the higher-order PNG and discuss the required survey level needed to test the consistency relation.
As future representative surveys, we consider Euclid-~\footnote{See http://www.euclid-ec.org} in optical and infrared bands, and the
Square Kilometre Array (SKA)-~\footnote{See http://www.skatelescope.org}
in radio wavelength. Both telescopes will perform ultimate galaxy surveys
with different characteristics.
The Euclid photometric galaxy survey ($15,000\,{\rm deg}^2$) will reach $z\approx 2.5$ and provides redshift information, which is highly advantageous to constrain $f_{\rm NL}$. On the other hand, the SKA continuum galaxy survey will cover a wider area of sky ($30,000\,{\rm deg}^2$) and 
a significant redshift depth ($z<5$), while the redshift information is not available. As we will show below, the wide and deep survey provided by SKA can constrain $\tau_{\rm NL}$ effectively.

\section{Angular power spectrum} 

In what follows, we compute the angular power spectrum for the halo density contrast, 
extending our previous work~\cite{Yamauchi:2014ioa} to include the effects of $\tau_{\rm NL}$ and $g_{\rm NL}$ (see also \cite{Kitching:2015fra,Takahashi:2015zqa}).
We first split whole samples into mass-divided subsamples for each redshift bin to apply the multitracer technique.
It can be shown that in the presence of $f_{\rm NL}$ and $\tau_{\rm NL}$
the halo-halo power spectrum between the $b$ th and $b^\prime$ th mass bins is expressed as~\cite{Smith:2010gx,Baumann:2012bc}
(see also \cite{Adhikari:2014xua})
\al{
	P^{(bb')}_{\rm hh}
		=&\biggl[
			\left( b_{\rm G}^{(b)}+f_{\rm NL}b_{\rm NG}^{(b)}\right)\left( b_{\rm G}^{(b')}+f_{\rm NL}b_{\rm NG}^{(b')}\right)
	\notag\\
	&\quad
			+\left(\frac{25}{36}\tau_{\rm NL}-f_{\rm NL}^2\right) b_{{\rm NG}}^{(b)}b_{{\rm NG}}^{(b')}
			\biggr] P_{\rm mm}
	\,,\label{eq:fNL-tauNL power spectrum}
}
where $b_{{\rm NG}}^{(b)}=\beta_f^{(b)}/{\cal M}D_+$,
$D_+(z)$ is the growth factor, and ${\cal M}(k)=2k^2T(k)/3\Omega_{\rm m,0}H_0^2$
with $T(k)$ being the matter transfer function normalized to unity at large scales~\cite{Eisenstein:1997ik}.
Throughout the paper, we use the expression $\beta_f^{(b)}=2\delta_{\rm c}(b_{\rm G}^{(b)}-1)$,
where $\delta_{\rm c}$ is the critical density for spherical collapse and we will take 
the $\delta_{\rm c}=1.46$ to fit the numerical simulation~\cite{Grossi:2009an}. 
We employ a fit to simulation for the mass function $\dd n/\dd M$ and 
the linear bias factor $b_{{\rm G}}^{(b)}$ given in \cite{Sheth:1999mn}.
If there is a nonvanishing $g_{\rm NL}$, we should take into account
the additional correction to the halo bias as 
$\Delta b_{\rm h}^{(b)}=g_{\rm NL}\beta_g^{(b)}/{\cal M}D_+$~\cite{Baumann:2012bc,Desjacques:2011mq,Ferraro:2014jba},
and we adopt the fitting function for $\beta_g^{(b)}$ to simulation~\cite{Smith:2011ub}.
When defining $f_{\rm NL}$, $\tau_{\rm NL}$, and $g_{\rm NL}$,
we have evaluated primordial fluctuations at present, though at the decoupling for the CMB convention,
suggesting that the observed nonlinear parameters have the relation~\cite{Camera:2014bwa}:
$f_{\rm NL}\approx 1.3f_{\rm NL}^{\rm CMB}$, $\tau_{\rm NL}\approx 1.3^2\tau_{\rm NL}^{\rm CMB}$,
and $g_{\rm NL}\approx 1.3g_{\rm NL}^{\rm CMB}$.
With Eq.~\eqref{eq:fNL-tauNL power spectrum}, the halo-halo angular power specta between $b$- and $b^\prime$-mass bins
in the $i$ th redshift bin is given by
\begin{widetext}
\al{
	C_{i(bb^\prime )}^{\rm hh}(\ell )
		=&\frac{2}{\pi}\int_0^\infty k^2\dd k\prod_{m=b,b^\prime}
			\biggl[
				\frac{1}{N_{i(m)}}\int_0^\infty\dd z_m\,j_\ell( k\chi ) \frac{\dd^2 V}{\dd z\dd\Omega}\int_0^\infty\dd M_m\frac{\dd n}{\dd M_m}S_{i(m)}
			\biggr]
		P^{(bb')}_{\rm hh}
	\,,\label{C_l def}
}
\end{widetext}
where $\dd^2 V/\dd z\dd\Omega =\chi^2 /H$, $\chi$ is the comoving distance,
and $N_{i(b)}=\int_0^\infty\dd z\frac{\dd^2 V}{\dd z\dd\Omega}\int_0^\infty\dd M\frac{\dd n}{\dd M}S_{i(b)}$ 
denotes the average density.
We have introduced $S_{i(b)}$ to represent the selection function.

To take advantage of the multitracer technique, we need to estimate the halo mass of each galaxy, which has to be inferred from available observables.
However, since estimates of the halo mass involve large uncertainties, a number of nuisance parameters should be included to model systematic errors.
We assume the mass-observable relation including uncertainties in the mass inference from available data.
In our treatment, the probability of assigning the estimated mass $M_{\rm est}$ to
the true mass $M$ is assumed to be given by log-normal distribution
with the variance $\sigma_{\ln M}(M,z)$ and the bias $\ln M_{\rm bias}(M,z)$~\cite{Lima:2004wn}.
With these, the selection function can be expressed as
$S_{i(b)}(M,z)=\Gamma_{(b)}\Theta (z-z_i)\Theta (z_{i+1}-z)
\frac{1}{2}\bigl[{\rm erfc}\left(x(M_{(b)};M)\right)-{\rm erfc}\left(x(M_{(b+1)};M)\right)\bigr]$,
where $x(M_{\rm est};M)=(\ln M_{\rm est}-\ln M-\ln M_{\rm bias})/\sqrt{2}\sigma_{\ln M}$, 
$\Gamma_{(b)}$ is the gray-body factor to denote the ratio between
halos and what we really observe for each mass bin, because we may not observe all galaxies associated with the underlying dark matter halos.
Hereafter we will introduce $14$ nuisance parameters in the variance and the bias 
to quantify the impact of possible residual systematic errors on the parameter estimation~\cite{Yamauchi:2014ioa,Oguri:2010vi}.

In order to understand the constraints on the non-Gaussian parameters, $f_{\rm NL}, \tau_{\rm NL}$ and $g_{\rm NL}$, it is useful to note that their effects on the halo-halo power spectrum have different redshift-, scale- and halo-mass dependences.
The correction to the bias roughly scales as 
$\beta_fz/k^2$ for $f_{\rm NL}$,
$\beta_f^2z^2/k^4$ for $\tau_{\rm NL}$, 
and $\beta_gz/k^2$ for $g_{\rm NL}$.
Because the correction to the bias from $f_{\rm NL}$ is identical to the one from 
$g_{\rm NL}=(\beta_f /\beta_g)f_{\rm NL}$,
$f_{\rm NL}$ and $g_{\rm NL}$ are degenerate for a single tracer case.
However, in the case with multiple tracers, the different halo-mass dependence of $\beta_f$ and $\beta_g$ would break their degeneracy~\cite{Ferraro:2014jba}.
On the other hand, because the bias correction due to $\tau_{\rm NL}$ has larger dependences on the scale and  redshift, its detectability would be enhanced for a survey with larger angular scale and higher redshift coverage.
Thus, we expect that, with the multitracer technique, wide and deep surveys will be powerful to probe higher-order PNG. 

\section{Results} 

To study the required survey level needed to test the consistency relation,
we proceed to the Fisher analysis. 
The Fisher matrix is defined by
$F_{\alpha\beta}=\sum_{\ell ,I,J}\frac{\pd\widehat C_I(\ell )}{\pd\theta^\alpha}\left({\rm Cov}^{-1}\right)_{IJ}\frac{\pd\widehat C_J(\ell )}{\pd\theta^\beta}$,
where the indices $I,J$ run over the redshift and mass bins $(i,b, b^\prime )$, 
$\theta^\alpha$ are free parameters to be determined by observations, and
$\widehat C_I(\ell )=C_{i(bb^\prime )}^{\rm hh}(\ell )+N_{i(b)}^{-1}\delta_{bb^\prime}^{\rm K}$ 
is the observed power spectrum including the shot noise contamination.
The marginalized expected $1\sigma$ error on $\theta^\alpha$ is estimated to be
$\sigma (\theta^\alpha )=\sqrt{(F^{-1})_{\alpha\alpha}}$.
We adopt the covariant matrix for multiple tracers whch are observed in different sky areas with
some overlap given in \cite{Yamauchi:2014ioa}.

Before showing the expected constraints, we need to specify the survey parameters. 
We consider the SKA radio continuum survey, the Euclid photometric survey, and their combination
as future representative surveys with significant high precisions.
SKA covers $30,000\,{\rm deg}^2$ out to $z=5$, though
there is only one redshift bin since the redshift information is not available.
Then we simply drop the redshift dependence in $\sigma_{\ln M}^{\rm SKA}$ and $\ln M_{\rm bias}^{\rm SKA}$.
To infer the halo masses, we consider five radio galaxy types such as star-forming galaxies, 
radio quite quasars, radio-loud AGN (FRI and FRII), and starbursts. 
Following \cite{Wilman:2008ew}, we will assign these galaxy types the following halo mass:
$\{M_{\rm SFG},M_{\rm RQQ},M_{\rm FRI},M_{\rm SB},M_{\rm FRII}\} =\{ 1,30,10^2,5\times 10^2,10^3\}$
in the unit of $10^{11}h^{-1}M_\odot$.
In order to have a plausible distribution for the halo mass associated with each population,
we introduce the five separating masses $M_{(1)}=0.9\times 10^{11}h^{-1}M_\odot$,
$M_{(2)}=\sqrt{M_{\rm SFG}M_{\rm RQQ}}$, $M_{(3)}=\sqrt{M_{\rm RQQ}M_{\rm FRI}}$,
$M_{(4)}=\sqrt{M_{\rm FRI}M_{\rm SB}}$, and $M_{(5)}=\sqrt{M_{\rm SB}M_{\rm FRII}}$.
With these, the five mass bins can be defined through $M_{(i)}<M<M_{(i+1)}$ $(i=1,2,3,4)$ and $M>M_{(5)}$.
To match the expected number density distribution of galaxies~\cite{Jarvis:2015tqa,Ferramacho:2014pua}, 
we adopt the gray-body factor 
$\Gamma_{(b)}^{\rm SKA1}=\{ 0.013,0.03,0.1,1,1\}$ and $\Gamma_{(b)}^{\rm SKA2}=\{0.2,0.4,1,1,1\}$, respectively.
As for Euclid, the covered area $15,000\,{\rm deg}^2$ and the redshift range $0.2<z<2.7$ are considered.
The redshift information is provided though photometric redshifts.
Galaxy samples are then split into five redshift bins with same interval $\Delta z=0.5$.
Since photometric surveys provide various galaxy properties such as luminosity, color, and stellar mass, 
which can be used to infer the halo mass, we can further split the galaxy samples according the estimated halo mass.
We consider five mass bins such that each mass bin of the same redshift bin has the same number density,
presumably because the tightest constraint is expected to be obtained when the shot noises for the mass bins 
become comparable.
For the flux cut, we adopt the minimum observed mass for each redshift bin, 
$M_{\rm est,min}/(10^{11}h^{-1}M_\odot )=0.7, 1,2,5,10$, and set $\Gamma_{(b)}^{\rm Euclid}=1$ instead.
For the combination of these surveys, the area of the overlap region is assumed to be $9,000\,{\rm deg}^2$
and we neglect the contributions from the derivative of the cross-correlations for simplicity.
In total we include $22=8$(SKA)$+14$(Euclid) nuisance parameters to model the systematic errors 
as well as three nonlinear parameters.
We choose $\sigma_{\ln M,0}^{\rm SKA}=1,\sigma_{\ln M,0}^{\rm Euclid}=0.3$, 
and zero for the other parameters as fiducial values.
To calculate the Fisher matrix, we use $2\leq\ell\leq 400$ for SKA and $3\leq\ell\leq 400$ for Euclid.
We hereafter focus on constraints on $f_{\rm NL}, \tau_{\rm NL}$, and $g_{\rm NL}$, marginalizing over the other parameters.
Our fiducial model is a standard $\Lambda$CDM cosmological model
with the parameters: 
$\Omega_{{\rm m},0}=0.318$, $\Omega_{{\rm b},0}=0.0495$, $\Omega_{\Lambda ,0} =0.6817$,
$w=-1$, $h=0.67$, $n_{\rm s}=0.9619$, $k_0=0.05{\rm Mpc}^{-1}$, and $\sigma_8=0.835$.

\begin{figure}[tbp]
\includegraphics[width=85mm]{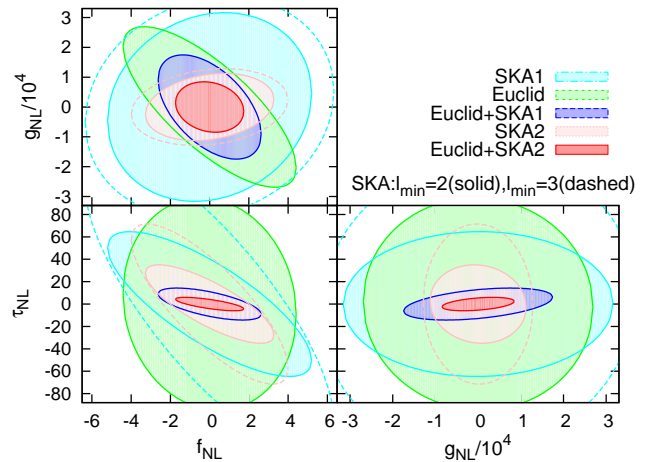}
\caption{
Forecast $1\sigma$ marginalized contours in $(f_{\rm NL},\tau_{\rm NL})$, $(g_{\rm NL}, \tau_{\rm NL})$,
and $(f_{\rm NL},g_{\rm NL})$ planes with the vanishing fiducial values of the nonlinear parameters.
To see the dependence on the minimal multipole we also plot the results for 
SKA with $\ell_{\rm min}^{\rm SKA}=3$ in the dashed line.
}
\label{fig:fNL-tauNL-gNL_Fisher}
\end{figure} 

In order to see the impact of the higher-order PNG on the parameter estimation, 
we first plot the $1\sigma$ expected marginalized contours in Fig.~\ref{fig:fNL-tauNL-gNL_Fisher}, in the case with vanishing nonlinear parameters.
Although the resultant constraints on $f_{\rm NL}$ are slightly weaker than the previous results~\cite{Yamauchi:2014ioa,Ferramacho:2014pua} where $\tau_{\rm NL}$ and $g_{\rm NL}$ were neglected,
the constraints from both SKA and Euclid are still significant, $\sigma (f_{\rm NL})={\cal O}(1)$.
Especially, the redshift information obtained from the photometric survey 
significantly improves the constraint on $f_{\rm NL}$~\cite{Yamauchi:2014ioa}.
As for $\tau_{\rm NL}$, SKA2 can reach $\sigma (\tau_{\rm NL})=23$, which is an improvement by a factor of $100$ compared with the Planck constraint.
In contrast, the constraint from Euclid, $\sigma (\tau_{\rm NL})=62$, 
is relatively weaker than one from SKA1, $\sigma (\tau_{\rm NL})=43$.
This is understood as follows. The bias correction from $\tau_{\rm NL}$ has stronger dependence on the wavelength and redshift than those from $f_{\rm NL}$ and $g_{\rm NL}$ as argued above.
Then wider sky coverage and deeper redshift are expected to be more advantageous to constrain $\tau_{\rm NL}$.
Although the redshift information is not available for SKA, the contributions from high-$z$ samples are significant to the effective bias.
To see the effect of sky coverage, we also plot the expected contours for SKA with $\ell_{\rm min}=3$ in the dashed line.
We find that $\ell_{\rm min}=3$ reduces the power of the SKA significantly and becomes comparable or weaker than Euclid, 
suggesting that the full-sky observation of the SKA is essential when we try to constrain the higher-order PNG. 
Similarly, the expected constraints on $g_{\rm NL}$ are given
as $\sigma (g_{\rm NL})=1.8\times 10^4$ (Euclid), $2.1\times 10^4$ (SKA1), and $7.4\times 10^3$ (SKA2), 
which are several tens of times severer than that obtained from Planck. 
These different behaviors suggest that SKA and Euclid turned out to be complementary probes of PNG and the joint analysis between SKA and Euclid are quite effective to confirm the PNG consistency inequality.

\begin{figure}[tbp]
\includegraphics[width=85mm]{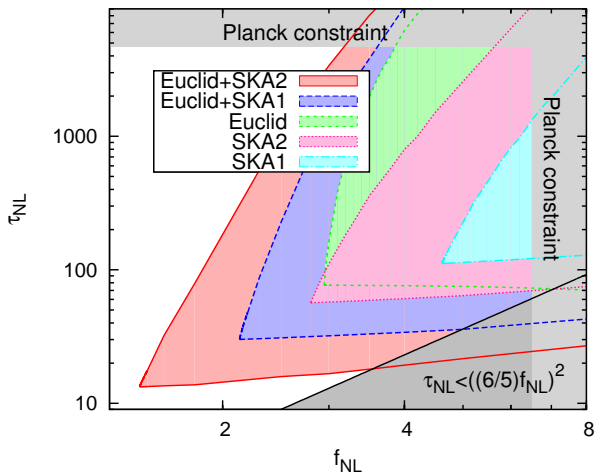}
\caption{
Parameter space to confirm the consistency relation, $f_{\rm NL}/\sigma (f_{\rm NL})\geq 1$ and
$\tau_{\rm NL}/\sigma (\tau_{\rm NL})\geq 1$.
For comparison, the inconsistent region and the constraints from Planck are also shown in gray color.
}
\label{fig:sig}
\end{figure} 

We then study the dependence of our forecast on the choice of the fiducial values, concentrating on $f_{\rm NL}$ and $\tau_{\rm NL}$.
For this purpose, we consider stochastic bias with values of $\tau_{\rm NL}$ which satisfies $\tau_{\rm NL} \geq ((6/5)f_{\rm NL})^2$.
In Fig.~\ref{fig:sig}, we show the region where both $f_{\rm NL}$ and $\tau_{\rm NL}$ are detected at $1\sigma$ significance, that is, $f_{\rm NL}/\sigma (f_{\rm NL})\geq 1$ and $\tau_{\rm NL}/\sigma (\tau_{\rm NL})\geq 1$ in $(f_{\rm NL},\tau_{\rm NL})$ plane for SKA1(2), Euclid, and their combinations.
When $\tau_{\rm NL}$ is close to the nonstochastic value,
there is little stochasticity and the tightest constraint on $f_{\rm NL}$ is obtained, as expected.
On the other hand, with increasing $\tau_{\rm NL}$, the constraining power on $f_{\rm NL}$ decreases, 
mainly because the correction from $\tau_{\rm NL}$ to the halo bias dominates that from $f_{\rm NL}$.
Hence, the relatively smaller $\tau_{\rm NL}$ is needed to detect $f_{\rm NL}$.
We find from the figure that even for SKA1 there is a wedge-shaped region where we can confirm the consistency inequality at more than $1\sigma$ level.
As is anticipated, the confirmation of the inequality becomes harder as decreasing $f_{\rm NL}$.
When $f_{\rm NL}$ is small, say $\lesssim 2.7$, the confirmation at the $\gtrsim 1\sigma$ level is rather
challenging for a single survey, even with the multitracer technique.
However, combining SKA2 and Euclid can break the degeneracy between $f_{\rm NL}$ and $\tau_{\rm NL}$,
and drastically improve the constraints to reach the confirmation of the consistency relation even for $f_{\rm NL}\approx 1.5$ and $\tau_{\rm NL}\approx 17$.
We roughly estimate the boundary of the viable region as 
$f_{\rm NL}\gtrsim 1$ and $10\lesssim\tau_{\rm NL}\lesssim 1.5(f_{\rm NL})^7$.

\section{Summary} 

To summarize, we have discussed the potential impact of the higher-order nonlinear parameters $\tau_{\rm NL}\,,g_{\rm NL}$,
and the required survey level needed to test the PNG consistency relation $\tau_{\rm NL}\geq ((6/5)f_{\rm NL})^2$
for future galaxy surveys such as SKA radio continuum and Euclid photometric surveys.
With the multitracer technique, 
the Fisher matrix analysis has revealed that 
deep survey and large sky coverage provided by SKA are advantageous to constrain $\tau_{\rm NL}$,
while the Euclid photometric survey has the strong constraining power for $f_{\rm NL}$ thanks to redshift information.
The information from both Euclid and SKA is quite essential to break the degeneracy between the nonlinear parameters.
Indeed the combination of SKA2 and Euclid can detect the consistency inequality in the wide parameter region at more than $1\sigma$ level, 
more specifically even for $f_{\rm NL}\approx 1.5$ and $\tau_{\rm NL}\approx 17$,
though for a single survey it is still hard to confirm when $f_{\rm NL}\lesssim 2.7$.

Our analysis also implies that the large-angle observations is quite essential to provide the constraints on the higher-order PNG. 
We should address the spectroscopic surveys conducted by both 
the SKA and Euclid. In this paper we conservatively assumed no redshift 
information for the SKA and relatively large redshift bin for the Euclid.
The redshift information for individual galaxies obtained from the spectropic surveys may allow the tomographic analysis 
to enhance the Fisher matrix due to the cross-correlations between the different redshift bins,
providing the improvement of the constraints.
We hope to come back to these issues in the near future.

When the constraints on nonlinear parameters, in particular $f_{\rm NL}$, are close to ${\cal O}(1)$, the general relativistic correction 
in the observed power spectrum might not be negligible~\cite{Camera:2014dia,Camera:2014sba,Fonseca:2015laa,Alonso:2015sfa}. 
We have simply neglected this effect expecting that we can subtract the effect from the observed power spectrum.

\acknowledgements
D.Y. is supported by Grant-in-Aid for JSPS Fellows (No.~259800). 
K.T. is supported by Grand-in-Aid from the Ministry of Education, Culture, Sports, and Science and Technology (MEXT) of Japan, No.~24340048,  No.~26610048 and No.~15H05896.


\end{document}